\pdfoutput=1
{\def\usepackage{ws-procs9x6}}

\documentclass[12pt,a4paper]{article}

\usepackage{amsmath,amssymb}
\usepackage[bookmarks=true,hyperfigures=true]{hyperref}
\usepackage{graphicx}
\usepackage[nosort]{cite}
\usepackage[bulletsep]{collref}

\providecommand{\hypersetup}[1]{}
\providecommand{\hyperref}[2][]{#2}
\providecommand{\texorpdfstring}[2]{#1}
\providecommand{\pdfbookmark}[3][]{}

\hypersetup{plainpages=false}
\hypersetup{pdfpagemode=UseOutlines}
\hypersetup{bookmarksnumbered=true}
\hypersetup{bookmarksopen=true}
\hypersetup{pdfstartview=FitH}
\hypersetup{colorlinks=false}
\hypersetup{citebordercolor={.6 .9 .6}}
\hypersetup{urlbordercolor={.7 .8 1}}
\hypersetup{linkbordercolor={1 .7 .7}}
\hypersetup{pdfborder={0 0 .5}}

\makeatletter
\let\@keywords\@empty
\let\@subject\@empty
\providecommand{\keywords}[1]{\gdef\@keywords{#1}}
\providecommand{\subject}[1]{\gdef\@subject{#1}}
\def\thetitle{\@title}
\def\theauthor{\@author}
\def\thesubject{\@subject}
\def\thedate{\@date}
\def\thekeywords{\@keywords}
\makeatother
\AtBeginDocument{
\hypersetup{pdftitle={\thetitle}}%
\hypersetup{pdfauthor={\theauthor}}%
\hypersetup{pdfsubject={\thesubject}}%
\hypersetup{pdfkeywords={\thekeywords}}%
}

\makeatletter
\newlength{\apb@width}
\newcommand{\autoparbox}[2][c]{\settowidth{\apb@width}{#2}\parbox[#1]{\apb@width}{#2}}
\newcommand{\includegraphicsbox}[2][]{\autoparbox{\includegraphics[#1]{#2}}}
\makeatother

\allowdisplaybreaks[3]

\numberwithin{equation}{section}

\usepackage[a4paper,text={450pt,650pt},centering]{geometry}

\usepackage[font=small,labelfont=bf,width=0.85\textwidth]{caption}

\let\oldbfseries=\bfseries
\let\oldmdseries=\mdseries
\let\oldnormalfont=\normalfont
\renewcommand{\bfseries}{\oldbfseries\boldmath}
\renewcommand{\mdseries}{\oldmdseries\unboldmath}
\renewcommand{\normalfont}{\oldnormalfont\unboldmath}

\DeclareMathSymbol{\Gamma}{\mathalpha}{letters}{"00}
\DeclareMathSymbol{\Delta}{\mathalpha}{letters}{"01}
\DeclareMathSymbol{\Theta}{\mathalpha}{letters}{"02}
\DeclareMathSymbol{\Lambda}{\mathalpha}{letters}{"03}
\DeclareMathSymbol{\Xi}{\mathalpha}{letters}{"04}
\DeclareMathSymbol{\Pi}{\mathalpha}{letters}{"05}
\DeclareMathSymbol{\Sigma}{\mathalpha}{letters}{"06}
\DeclareMathSymbol{\Upsilon}{\mathalpha}{letters}{"07}
\DeclareMathSymbol{\Phi}{\mathalpha}{letters}{"08}
\DeclareMathSymbol{\Psi}{\mathalpha}{letters}{"09}
\DeclareMathSymbol{\Omega}{\mathalpha}{letters}{"0A}

\newcommand{\gen}[1]{\mathfrak{#1}}
\newcommand{\GG}{\mathfrak{J}}

\newcommand{\geny}[1]{\mathfrak{\widehat{#1}}}
\newcommand{\alg}[1]{\mathfrak{#1}}
\newcommand{\grp}[1]{\mathrm{#1}}
\newcommand{\yangian}[1]{\mathrm{Y}[#1]}
\newcommand{\superN}{\mathcal{N}}
\newcommand{\sym}{$\superN=4$ SYM}
\newcommand{\scs}{$\superN=6$ SCS}
\newcommand{\psu}{\ensuremath{\alg{psu}(2,2|4)}}
\newcommand{\osp}{\ensuremath{\alg{osp}(6|4)}}
\newcommand{\cp}{\grp{\Complex P}}
\newcommand{\amp}{A}
\newcommand{\lambdaYM}{\lambda\indups{YM}}

\newcommand{\order}[1]{\mathcal{O}(#1)}
\newcommand{\pardel}[1]{\frac{\partial}{\partial#1}}
\newcommand{\sgn}{\mathop{\mathrm{sign}}}
\newcommand{\transpose}{^\mathrm{T}}

\newcommand{\Complex}{\mathbb{C}}

\newcommand{\ff}{f\kern-5pt f}

\newcommand{\dd}{d}
\newcommand{\eps}{\varepsilon}

\newcommand{\deltad}[1]{\delta^{#1}}

\ifx\genfrac\sdflkaj\else\fi
\newcommand{\sfrac}[2]{{\textstyle\frac{#1}{#2}}}
\newcommand{\half}{\sfrac{1}{2}}
\newcommand{\ihalf}{\sfrac{i}{2}}

\newcommand{\indup}[1]{_{\mathrm{#1}}}
\newcommand{\indups}[1]{_{\mathrm{\scriptscriptstyle #1}}}
\newcommand{\supup}[1]{^{\mathrm{#1}}}

\newcommand{\twi}[1]{\mathcal{#1}}
\newcommand{\twind}[1]{\mathcal{#1}}

\newcommand{\lrbrk}[1]{\left(#1\right)}
\newcommand{\bigbrk}[1]{\bigl(#1\bigr)}
\newcommand{\biggbrk}[1]{\biggl(#1\biggr)}

\newcommand{\comm}[2]{[#1,#2]}

\newcommand{\sprod}[2]{\langle#1,#2\rangle}

\newcommand{\cprod}[2]{[#1,#2]}

\newcommand{\levind}[1]{[#1]}

\newcommand{\nn}{\nonumber}
\newcommand{\nln}{\nonumber\\}

\newcommand{\beq}{\begin{equation}}
\newcommand{\eeq}{\end{equation}}

\def\[{\begin{equation}}
\def\]{\end{equation}}
\def\<{\begin{eqnarray}}
\def\>{\end{eqnarray}}

\makeatletter
\def\mr@ignsp#1 {\ifx\:#1\@empty\else #1\expandafter\mr@ignsp\fi}%
\newcommand{\multiref}[1]{\begingroup
\xdef\mr@no@sparg{\expandafter\mr@ignsp#1 \: }%
\def\mr@comma{}%
\@for\mr@refs:=\mr@no@sparg\do{\mr@comma\def\mr@comma{,}\ref{\mr@refs}}%
\endgroup}
\makeatother

\renewcommand{\eqref}[1]{(\multiref{#1})}

\newcommand{\namedref}[2]{\hyperref[#2]{#1~\ref*{#2}}}
\newcommand{\Secref}[1]{\namedref{Section}{#1}}
\newcommand{\secref}[1]{\namedref{Section}{#1}}

\newcommand{\Figref}[1]{\namedref{Figure}{#1}}
\newcommand{\figref}[1]{\namedref{Figure}{#1}}

\providecommand{\href}[2]{#2}
\newcommand{\arxivlink}[1]{\href{http://arxiv.org/abs/#1}{arxiv:#1}}

\title{Exact Superconformal and Yangian Symmetry\texorpdfstring{\\}{ }of Scattering Amplitudes}
\author{Till Bargheer\texorpdfstring{$^a$}{}, Niklas Beisert\texorpdfstring{$^b$}{}, Florian Loebbert\texorpdfstring{$^c$}{}}

\begin{document}

\pdfbookmark[1]{Title Page}{title}

\thispagestyle{empty}
\begin{flushright}\footnotesize
\texttt{\arxivlink{1104.0700}}\\
\texttt{AEI-2011-016}\\%
\texttt{LPT ENS-11/12}\\%
\href{http://www.fysast.uu.se/teorfys/en/content/exact-superconformal-and-yangian-symmetry-scattering-amplitudes}{\texttt{UUITP-11/11}}\\%
\end{flushright}
\vspace{1cm}

\begin{center}%
\begingroup\Large\bfseries\thetitle\par\endgroup
\vspace{1cm}%

\begingroup\scshape\theauthor\par\endgroup
\vspace{5mm}%

\begingroup\itshape
$^a$
Department of Physics and Astronomy\\
Uppsala University\\
SE-751 08 Uppsala, Sweden
\vspace{3mm}

$^b$
Max-Planck-Institut f\"ur Gravitationsphysik\\
Albert-Einstein-Institut\\
Am M\"uhlenberg 1, 14476 Potsdam, Germany
\vspace{3mm}

$^c$
Laboratoire de Physique Th\'eorique\\
\'Ecole Normale Sup\'erieure\\
24 Rue Lhomond, 75005 Paris, France
\par\endgroup
\vspace{5mm}

\begingroup\ttfamily
\texttt{till.bargheer@physics.uu.se, nbeisert@aei.mpg.de, loebbert@lpt.ens.fr}
\par\endgroup

\vspace{1cm}

\textbf{Abstract}\vspace{7mm}

\begin{minipage}{12.7cm}
We review recent progress in the understanding of symmetries 
for scattering amplitudes in $\superN=4$ superconformal Yang--Mills theory. 
It is summarized how the superficial breaking of superconformal symmetry 
by collinear anomalies and the renormalization process can be cured 
at tree and loop level. 
This is achieved by correcting the representation of the superconformal 
group on amplitudes. Moreover, we comment on the Yangian symmetry 
of scattering amplitudes and how it inherits these correction terms 
from the ordinary Lie algebra symmetry. Invariants under this algebra 
and their relation to the Gra{\ss}mannian generating function
for scattering amplitudes are discussed. 
Finally, parallel developments 
in $\superN=6$ superconformal Chern--Simons theory are summarized.
This article is an invited review for a special issue 
of Journal of Physics A devoted to \emph{Scattering Amplitudes in Gauge Theories}.
\end{minipage}

\end{center}

\newpage

\setcounter{tocdepth}{2}
\hrule height 0.75pt
\pdfbookmark[1]{\contentsname}{contents}
\tableofcontents
\vspace{0.8cm}
\hrule height 0.75pt
\vspace{1cm}

\setcounter{tocdepth}{2}

\section{Introduction}

An efficient unitarity-based construction of the scattering matrix%
\footnote{See \cite{Bern:2011qt,Carrasco:2011hw} within this special
issue.}
in a quantum field theory relies heavily on
the concepts of locality, analyticity and symmetry.
Symmetries are tremendously important
because they strongly constrain the permissible building blocks
from the start and guide reliably towards the desired final result.
This is especially true for highly symmetric theories such as $\superN=4$ super
Yang--Mills theory (SYM),
which is
believed to be integrable in the planar limit,
cf.\ \cite{Beisert:2010jr}. During the last
few years, remarkable structures in this theory's scattering amplitudes have been
discovered. Most
notably, planar amplitudes display a hidden `dual' superconformal symmetry
\cite{Drummond:2006rz,Drummond:2008vq,Brandhuber:2008pf},%
\footnote{See also \cite{Drummond:2011..,Henn:2011xk} within this special issue.}
which together with the ordinary
superconformal symmetry combines into Yangian symmetry
\cite{Drummond:2009fd}. The latter is a typical feature of integrable
models (see \cite{Bernard:1993ya,MacKay:2004tc} for reviews), and was observed earlier in the
spectral problem of the theory \cite{Dolan:2003uh}.
Commonly, the spectrum and dynamics of integrable models are strongly
constrained or even completely determined by the extended symmetry.
Conceivably, this is also the case for \sym\ amplitudes. For exploiting the
constraints, a thorough understanding of the symmetries is indispensable.
Here, we review the status of superconformal and Yangian symmetry for \sym\
scattering amplitudes. We also comment on parallel developments in
three-dimensional $\superN=6$ super Chern--Simons (SCS)
theory.

\medskip
\noindent
Scattering amplitudes in conformal field theories show infrared
divergences. Their regularization by means of a mass scale
superficially
breaks conventional conformal symmetry. On the other hand,
superconformal symmetry in \sym\ is expected to be exact also at the quantum level.
Is it possible to reconcile the symmetry with
a non-vanishing regulator that is required for a consistent
formulation of scattering amplitudes?
Can the symmetry breaking be assessed quantitatively, or, even better,
can the broken symmetry be restored in a modified way?
Interestingly, a careful study reveals that superconformal symmetry is broken 
already at tree level
\cite{Cachazo:2004dr,Britto:2004nj}. Namely, acting with a free generator on a
tree-level amplitude produces residual contributions whenever two external
legs become collinear.
Exact superconformal invariance can be restored by introducing a non-linear
correction to the generator that cancels the residual term \cite{Bargheer:2009qu}. Importantly, only the
\sym\ scattering matrix as a whole is exactly invariant, its individual
entries (the amplitudes) are not.
While only contributing to singular momentum configurations at tree level,
collinear residues become inevitable at higher orders, where
loop momenta are integrated over. At one-loop order, superconformal symmetry can again be
restored by further generator corrections, which cures residual
contributions
both from collinear terms and from infrared regularization
\cite{Sever:2009aa,Beisert:2010gn}.

The corrections for the superconformal generators
straightforwardly carry over to the Yangian symmetry of scattering
amplitudes.
The formally very simple Gra{\ss}mannian function of
\cite{ArkaniHamed:2009dn} generates invariants
of the free (uncorrected) Yangian \cite{Drummond:2010qh}.%
\footnote{The `invariants' generated by the Gra{\ss}mannian function are
not \emph{exact} invariants
-- they are only invariant under the free
(undeformed) symmetry up to residual contributions at collinear momenta.}
In fact, it is believed to generate \emph{all} free
Yangian invariants \cite{Drummond:2010uq,Korchemsky:2010ut}.
Scattering amplitudes are linear combinations of
these invariants satisfying physicality requirements such as correct
collinear limits or cancellation of unphysical poles
\cite{Korchemsky:2009hm}. Free Yangian symmetry alone is insufficient for
fixing the right linear combination. The missing piece is provided by the
generator corrections: It appears that they single out the physical linear
combination as the unique \emph{exact} invariant
\cite{Bargheer:2009qu}, thus paving the way for an algebraic
determination of loop amplitudes.

Compared to \sym, much less is known about scattering amplitudes in its
three-dimensional cousin, \scs\ theory
\cite{Bagger:2006sk,Gustavsson:2007vu,Bagger:2007jr,Bagger:2007vi,Gustavsson:2008bf,Bagger:2008se,Aharony:2008ug}
(or `ABJM' named after the authors of \cite{Aharony:2008ug}).
Both theories are surprisingly similar, and indeed, 
counterparts to some of the most important symmetry
structures known from \sym\ amplitudes have been found in \scs\ during the last
year. In particular, there is compelling evidence for Yangian and dual
superconformal symmetry \cite{Bargheer:2010hn,Huang:2010qy,Gang:2010gy}.
Nevertheless, several fundamental questions 
regarding symmetries of the S-matrix in \scs\ remain to be answered.

\medskip
\noindent
This work is structured as follows: We review how exact superconformal
symmetry is restored at tree level in \secref{sec:exact},
and we also comment on the extension to loops. 
In \secref{sec:yangsym}, we briefly recapitulate Yangian symmetry in the
context of \sym\
scattering amplitudes, and remark on the corrections to Yangian generators.
The Gra{\ss}mannian generating function for tree-level invariants
and the implications of the generator corrections
for these invariants are discussed in \secref{sec:invagra}.
\secref{sec:abjm} summarizes what is known about the symmetry structures of
scattering amplitudes in \scs\ theory.

\section{Exact Superconformal Symmetry}
\label{sec:exact}

Maximally supersymmetric Yang--Mills theory
is a four-dimensional conformal field theory. 
It is therefore natural to assume that 
its S-matrix is exactly invariant 
under the superconformal algebra $\alg{psu}(2,2|4)$.
Invariance of the S-matrix is, however, 
not straight-forward.
First of all, the presence of massless particles 
inevitably leads to infrared divergences
in scattering amplitudes at loop level. 
A regulator for the divergences breaks 
conformal symmetry, e.g.\ 
by moving away from four spacetime dimensions or
by introduction of a mass scale.
Only after the regulator is removed,
we can hope for a restoration of conformal symmetry,
but a priori there is no guarantee. 
In case of success, the procedure 
will most likely have deformed and thus obscured the action
of the symmetry on the (renormalized) S-matrix.%
\footnote{There is an alternative treatment of
(dual) conformal symmetries for \sym\ on the Coulomb branch.
This is discussed in detail in \cite{Henn:2011xk}
within this special issue. We will not comment on it here.}

\subsection{Tree Level}
\label{sec:exacttree}

In fact, the situation is even more subtle than this. 
Let us consider a color-ordered MHV amplitude at tree level,%
\footnote{For a more detailed introduction to the formalism,
see \cite{Dixon:2011..,Brandhuber:2011ke,Elvang:2010xn,Bern:2011qt} within this special issue.}
\[\label{eq:MHV}
A\supup{MHV}_n
=
\frac{\delta^4(P)\,\delta^8(Q)}{\prod_{k=1}^n\sprod{k}{k+1}}\,,
\qquad
\begin{aligned}
P^{b\dot a}&=\textstyle\sum_{k=1}^n p_k^{b\dot a},
&
p^{b\dot a}_k&=\lambda_k^b\tilde\lambda_k^{\dot a},
\\
Q^{b A}&=\textstyle\sum_{k=1}^n q_k^{b A},
&
q^{b A}_k&=\lambda_k^b\eta_k^{A}.
\end{aligned}
\]
We use the spinor helicity formalism
to encode particle momenta and flavors:
The $k$-th particle is described by the bosonic spinor
$\lambda_k\in\Complex^{2}$
with complex conjugate
$\tilde\lambda_k=\pm\bar\lambda_k$
(the sign determines the sign of the energy)
and the fermionic spinor 
$\eta_k\in\Complex^{0|4}$.
The two mutually conjugate Lorentz-invariant 
spinor products are denoted by 
\[
\sprod{\lambda}{\mu}=\varepsilon_{ac}\lambda^a\mu^c,
\qquad
\cprod{\tilde\lambda}{\tilde\mu}=\varepsilon_{\dot a\dot c}\tilde\lambda^{\dot a}\tilde\mu^{\dot c}.
\]
Now we act on the MHV amplitude with a free superconformal generator 
\[\label{eq:Sbar0}
\gen{\bar S}^B_{\dot a}=
\sum_{k=1}^n \eta_k^B\frac{\partial}{\partial \tilde\lambda_k^{\dot a}}\,.
\]
Superficially, the derivative acts only on $P$ in $\delta^4(P)$ and 
produces a factor of $Q$. 
The fermionic delta function $\delta^8(Q)$ makes the result vanish,
i.e., the MHV amplitude is invariant under $\gen{\bar S}$.

Interestingly, this is not the full story:
There is a subtle contribution when the derivative w.r.t.\ $\tilde\lambda_k$
hits a pole of the (otherwise) holomorphic denominator in \eqref{eq:MHV}
\cite{Cachazo:2004dr,Britto:2004nj}.
The so-called holomorphic anomaly for the spinor product reads
($E(\lambda)$ is the energy associated to the spinor $\lambda,\tilde\lambda$)
\[
\label{eq:anomaly}
\frac{\partial}{\partial \tilde\lambda^{\dot a}}\,
\frac{1}{\sprod{\lambda}{\mu}}
=2\pi \sgn\bigbrk{E(\lambda) E(\mu)} \varepsilon_{\dot a\dot c}
\tilde\mu^{\dot c}
\delta^2\bigbrk{\sprod{\lambda}{\mu}}.
\]
In the MHV amplitude this yields distributional contributions
supported on kinematical configurations 
where $\sprod{k}{k+1}=\cprod{k}{k+1}=0$.
In other words, this means that invariance of the S-matrix under free superconformal transformations
is violated where the momenta of two adjacent particles are collinear. 
The terms that break invariance can be summarized as follows
\cite{Bargheer:2009qu}
\[
\label{eq:violation}
\gen{\bar S}^B_{\dot a}
A_n=-
\sum_{k=1}^n
\int d^{4|4}\Lambda\,
\bar S^{B}_{\dot a}(k,k+1,\bar\Lambda)\,
A_{n-1}(1,\ldots,k-1,\Lambda,k+2,\ldots,n).
\]
Here $\Lambda:=(\lambda,\tilde\lambda,\eta)$
and $\bar\Lambda:=(\lambda,-\tilde\lambda,-\eta)$.
The integral over $\Lambda$ sums over all flavors and light-like momenta
in a Lorentz-invariant fashion.
The kernel of superconformal violation reads
\begin{align}
\label{eq:kernel}
\bar S^B_{\dot a}(1,2,3)
=&
-2\varepsilon_{\dot a\dot c}\tilde\lambda_3^{\dot c}
\int d^{4|4}\Lambda'\,\delta^4(\lambda')\,
\eta'^B
\int_{0}^{\pi/2} d\alpha\int_{0}^{2\pi} d\varphi\int_{0}^{2\pi} d\vartheta\,e^{i\varphi+i\vartheta}
\nln &
\qquad\cdot
\delta^{4|4}(e^{-i\varphi}\bar\Lambda_3\sin\alpha+e^{i\vartheta}\bar\Lambda'\cos\alpha-\Lambda_1)
\nln &
\qquad\cdot
\delta^{4|4}(e^{-i\vartheta}\bar\Lambda_3\cos\alpha-e^{i\varphi}\bar\Lambda'\sin\alpha-\Lambda_2)
+\mbox{2 cyclic images},
\end{align}
where $z\Lambda:=(z\lambda,\bar z\tilde\lambda,\bar z\eta)$ for $z\in\Complex$.
The delta function $\delta^4(\lambda')$ enforces collinearity
of all three momenta,
and $\sin^2\alpha,\cos^2\alpha$ represent the momentum fractions
for particles $1,2$, respectively,
in terms of particle $3$.

\begin{figure}
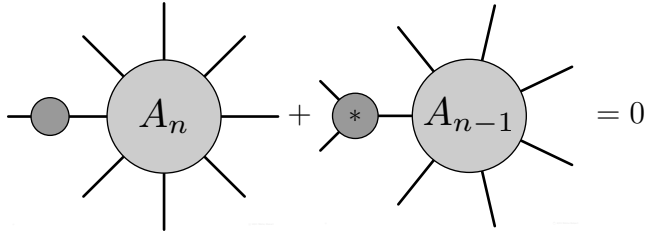
\centering
$\includegraphicsbox{FigInvTree1.mps}+\includegraphicsbox{FigInvTree2.mps}=0$
\caption{Deformed superconformal invariance relation for tree amplitudes.}
\label{fig:exactinvariance}
\end{figure}

Importantly, the free superconformal violation of $A_n$ in \eqref{eq:violation}
is expressed through another tree amplitude $A_{n-1}$. 
We introduce an operator $\gen{\bar S}_+$ 
which attaches the kernel $\bar S$ to an amplitude function 
as in minus the r.h.s.\ of \eqref{eq:violation},
cf.\ \figref{fig:exactinvariance}.
Then we can write
\[
\gen{\bar S}A_n+\gen{\bar S}_+A_{n-1}=0.
\]
Although individual scattering amplitudes $A_n$ with a fixed number
of external legs are not exactly conformally invariant, 
the full S-matrix 
(representing the generating functional for all amplitudes) 
is invariant under the
deformed superconformal generator
$\gen{\bar S}+\gen{\bar S}_+$.

Until now we have only discussed MHV amplitudes \eqref{eq:MHV}.
Luckily, all of the above applies to generic N$^k$MHV tree amplitudes as well.
The reason lies in the universality of collinear behavior:
A scattering amplitude $A_n$ diverges
in the vicinity of collinear momentum configurations \cite{Mangano:1987xk,Bern:1994zx,Kosower:1999xi}.
The pole is given by the amplitude $A_{n-1}$
with one fewer leg times a universal splitting function.
Superconformal generators yield a distributional term \eqref{eq:violation} 
at these poles. The kernel $\bar S$ is essentially 
the superconformal variation of the splitting function.
Note that the splitting function is more or less equivalent to
the three-point function $A_3$ which 
cannot exist in Minkowski signature. 
In split signature, however, one can derive the kernel as the variation
of the three-point function, $\bar S=-\half\gen{\bar{S}}A_3$
\cite{Beisert:2010gn}. 

Similar considerations hold for the conjugate superconformal generator $\gen{S}$
and the bosonic conformal generator $\gen{K}$.
The latter in fact receives a further correction $\gen{K}_{++}$
that maps one leg to three.
On the other hand, all super-Poincar\'e generators $\gen{P},\gen{Q},\gen{\bar Q},\gen{L},\gen{\bar L},\gen{R}$
as well as the dilatation generator $\gen{D}$ are manifest symmetries
of the tree level S-matrix.

\subsection{Further Considerations}
\label{sec:furthcon}

All in all this shows that the complete tree-level S-matrix 
is indeed exactly conformally invariant,
but only under the interacting superconformal generator
$\gen{\bar S}+\gen{\bar S}_+$.
This observation calls for a few clarifications
to be discussed in the following.

\emph{Does the above apply to gauge groups other than $\grp{SU}(N\indup{c})$?}
The answer is affirmative:
The kernel in \eqref{eq:kernel} must be complemented 
with the structure constants for the gauge group.
The free superconformal variation in \eqref{eq:violation}
generalizes canonically \cite{Bargheer:2009qu}.

\emph{Does it mean that superconformal symmetry is anomalous at tree level?}
In quantum field theory an anomaly refers 
to a violation of symmetry which cannot be repaired,
at least not by a local deformation. 
Here superconformal symmetry becomes exact when the deformation is included.
Moreover the deformation has no poles or cuts, it is local. 
It is not an anomaly, but rather a careful treatment of a non-manifest symmetry.%
\footnote{We thank H.~Nicolai for discussions of this issue.}

\emph{Does the deformation alter the $\alg{psu}(2,2|4)$ superconformal algebra?}
It is a proper representation,
albeit of a somewhat bigger algebra \cite{Bargheer:2009qu}:
First of all, the anticommutator of 
the superconformal generator $\gen{S}$ and its conjugate $\gen{\bar S}$
consistently defines the deformed conformal generator $\gen{K}$.
The only subtlety is in the anticommutators between two $\gen{S}$'s 
or two $\gen{\bar S}$'s:
They ought to vanish for $\alg{psu}(2,2|4)$, but they do not.
Instead they represent a gauge variation 
which transforms a covariant field $X$ according to $X\mapsto \comm{G}{X}$. 
Here the gauge variation parameter $G$ is actually a field itself, namely
the zero mode of the scalar field. 
Such a deformation of the algebra is not harmful
because it vanishes for all physically meaningful, i.e.\ gauge invariant, observables. 
In fact, it is very common in gauge theories 
that symmetry algebras are deformed by gauge variations, 
e.g.\ the supersymmetry algebra for gauge theories with extended supersymmetry.

\emph{Is there a physical reason for the deformation? What does it mean?}
Notice that the violation of free superconformal symmetry
occurs at collinear momentum configurations.
This points at the problems encountered in scattering theory 
for a model without a mass gap (in particular for a CFT),
see also \cite{Mason:2009sa}.
Scattering amplitudes require a notion of asymptotic particles
which do not interact further.
However, nothing prevents a massless particle from decaying into 
two or more massless particles at any time. 
Lorentz invariance implies that these particles necessarily 
have strictly collinear momenta. 
In physical terms such asymptotic decays 
have no implications because a detector
would merely measure the total deposited momentum and energy
of all particles emitted in a specific direction.
In other words, the Fock space for massless asymptotic particles 
is bigger than necessary. 
The physical space must be supplemented with an equivalence relation 
to factor out particle configurations with collinear momenta.
It is reasonable to relate the deformed representation with this issue. 
Presumably, the deformation makes superconformal transformations 
compatible with the structure of representatives of
the equivalence relation used for scattering amplitudes.%
\footnote{We thank D.~Skinner for discussions of this issue.} 

\emph{Is there a relation to the deformations for superconformal representations on
local operators?}
It is very analogous, and the same structures \cite{Beisert:2003ys} are observed,
cf.\ \cite{Beisert:2010jq}; quite likely it is equivalent to some extent. 
There are, however, important differences. 
For local operators the representation must deal with UV divergences. 
These are absent for scattering amplitudes leading to simplifications. 
For instance, the free super-Poincar\'e representation is undeformed
for amplitudes whereas it requires non-trivial deformations for local operators.
On the other hand, scattering amplitudes have IR divergences which are
absent for local operators. 

\begin{figure}
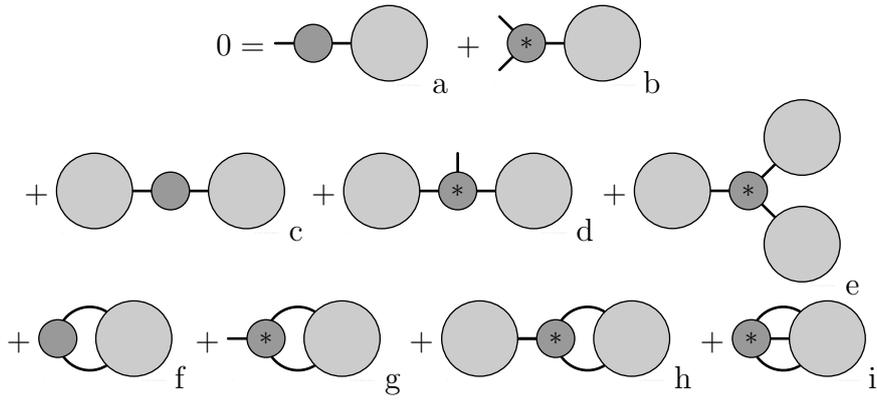
\centering
$0=\includegraphicsbox{FigInv1.mps}\indup{\textstyle a}
+\includegraphicsbox{FigInv2.mps}\indup{\textstyle b}$\\
$\mathord{}+\includegraphicsbox{FigInv3.mps}\indup{\textstyle c}
+\includegraphicsbox{FigInv4.mps}\indup{\textstyle d}
+\includegraphicsbox{FigInv5.mps}\indup{\textstyle e}$\\
$\mathord{}+\includegraphicsbox{FigInv6.mps}\indup{\textstyle f}
+\includegraphicsbox{FigInv7.mps}\indup{\textstyle g}
+\includegraphicsbox{FigInv8.mps}\indup{\textstyle h}
+\includegraphicsbox{FigInv9.mps}\indup{\textstyle i}$
\caption{General superconformal invariance relation (qualitatively). 
Terms a--b correspond to \protect\figref{fig:exactinvariance},
terms c--e are needed for factorized amplitudes,
and terms f--i are needed for loops.
A big circle represents a connected scattering amplitude. 
The small circle represents a free conformal generator (empty)
or the three-point kernel (starred).}
\label{fig:fullinvariance}
\end{figure}

\emph{Do multi-particle poles introduce further violations of free superconformal symmetry?}
Yes and no. The holomorphic anomaly produces a codimension-two distribution \eqref{eq:anomaly}. 
This matches with the codimension $D-2$ of collinear configurations 
of two massless particles in $D=4$ Minkowski space. 
Multi-particle poles are always codimension-one, 
thus the free superconformal generators 
do not yield distributional terms \cite{Bargheer:2009qu}.
This exhausts all singularities at tree level,
nevertheless one has to be careful \cite{Sever:2009aa,Beisert:2010gn}:
Multi-particle poles originate from Feynman propagators $1/(p^2\pm i\epsilon)$.
The principal part $1/p^2$ is harmless as explained above, 
but the on-shell contribution $\pm i\pi\delta(p^2)$
requires further deformations. 
The holomorphic anomaly also appears when an internal 
momentum becomes collinear with an external one
or even if two internal momenta become collinear. 
In a graphical representation where \eqref{eq:violation} (\figref{fig:exactinvariance})
is given by \figref{fig:fullinvariance}a,b,
the additional terms take the form of \figref{fig:fullinvariance}d,e \cite{Sever:2009aa}.
This completes the analysis at tree level.

\subsection{Loop Level}
\label{sec:exloop}

At tree level it is easy to ignore the distributional terms \eqref{eq:violation}
which break invariance under free superconformal transformations.
For generic configurations of the external momenta,
none of the internal or external momenta are collinear.
Consequently, the free superconformal generators annihilate 
the scattering amplitude.
At loop level, the situation is different.
Within the loop integrals some internal momenta inevitably 
become collinear with others.
Thus for generic configurations of the external momenta,
invariance under free superconformal transformations is broken.

To understand superconformal transformations at loop level 
it is important to quantify the violation terms.
Unfortunately, loop integrals are off-shell, 
and we cannot immediately address superconformal transformations
using the framework outlined above. 
This problem is circumvented by considering (generalized) unitarity cuts
\cite{Korchemsky:2009hm,Sever:2009aa}
which are expressed through on-shell amplitudes at lower loop orders.
Superconformal transformations of cuts 
can thus be obtained recursively
through the transformations at tree level.
At the end we will have to lift the transformation rule
from a cut integral to a full loop integral.

\smallskip

\begin{figure}
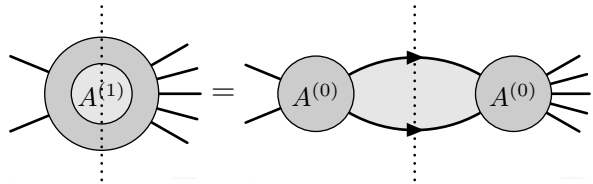
\centering 
$\includegraphicsbox{FigCutA1.mps}=
\includegraphicsbox{FigCutA0A0.mps}$
\caption{Essential contribution to the one-loop unitarity cut.}
\label{fig:cut}
\end{figure}

\begin{figure}
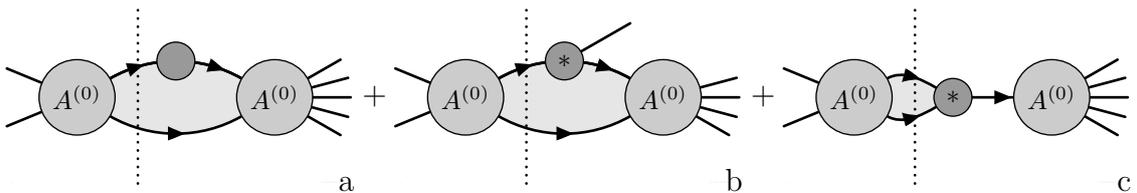
\centering
$\includegraphicsbox{FigCutAnom1.mps}_{\makebox[0pt][r]{a}}+
\includegraphicsbox{FigCutAnom2.mps}_{\makebox[0pt][r]{b}}+
\includegraphicsbox{FigCutAnom3.mps}_{\makebox[0pt][r]{c}}$
\caption{Essential contributions
to the one-loop superconformal variation.}
\label{fig:oneloopinvariance}
\end{figure}

Let us thus consider a simple one-loop unitarity cut \cite{Beisert:2010gn}.
The essential contribution is written as an on-shell integral over two tree-level amplitudes, 
see \figref{fig:cut} (we discard various other ways to cut the amplitude).
Now we can substitute the relevant superconformal transformation rule 
for tree amplitudes from \figref{fig:fullinvariance}. 
Discarding the correction terms already present at tree level, 
there are three new terms due to the superconformal generators
acting on internal legs, see \figref{fig:oneloopinvariance}.
Let us briefly discuss these terms. 

The term in \figref{fig:oneloopinvariance}a essentially
represents a superconformal transformation of an internal on-shell propagator. 
Propagators are superconformal invariants, 
so this contribution ought to be zero. 
Due to the appearance of IR divergences in the integrals, however,
we have to work with regularized propagators which violate invariance 
by a small amount. 
In this case, divergences only appear when one of the two subamplitudes 
has four legs, and when there is no momentum transfer between the two pairs of legs. 
In dimensional regularization, for example,
the (planar) correction to the
generator $\gen{D}$ of conformal rescalings reads 
\[\label{eq:corrD1}
\gen{D}(\lambdaYM)=\gen{D}^{(0)}+ \Gamma(\lambdaYM,\epsilon) \sum_{k=1}^n \gen{D}^{(1)}_{k,k+1},
\qquad
\gen{D}^{(1)}_{k,k+1}=-\frac{1}{2\epsilon} \left(\frac{(p_{k}+p_{k+1})^2}{-\mu^2}\right)^{-\epsilon}.
\]
For $\gen{D}$ this is the only correction at loop level,
and the coefficient in front 
\begin{equation}\label{eq:frontcoeff}
\Gamma(\lambdaYM,\epsilon)=\Gamma_{\mathrm{cusp}}(\lambdaYM)+\epsilon\Gamma_{\mathrm{coll}}(\lambdaYM)+\order{\epsilon^2}
\end{equation}
includes the cusp dimension $\Gamma_\mathrm{cusp}(\lambdaYM)=\lambdaYM/4\pi^2+\order{\lambdaYM^2}$ 
as well as the collinear dimension $\Gamma_\mathrm{coll}(\lambdaYM)=\order{\lambdaYM^2}$.
The other superconformal generators $\gen{S}, \gen{\bar S},\gen{K}$
receive analogous IR corrections, but they also receive corrections
due to the collinear behavior discussed in \secref{sec:exacttree}.

The term in \figref{fig:oneloopinvariance}b quantifies
the effect of superconformal transformations when
internal and external legs become collinear. 
It has the particularly nice property that the momenta running in the triangular
loop are all fixed by the on-shell conditions. 
Hence there is no integral to be performed and the result is rational. 
Removing the cut can be achieved simply by inserting an appropriate logarithm.
These terms take a lengthier form.

The last term in \figref{fig:oneloopinvariance}c is rather strange.
The kernel forces the two momenta across the cut to be collinear. 
Consequently, the subamplitude on the other side is evaluated at two collinear external momenta,
i.e.\ right on the pole. 
The problem of defining the result is directly related
to the ambiguity in defining the loop correction to the splitting function,
see \cite{Bern:1994zx,Kosower:1999xi}.
A justifiable resolution to the problem is to discard this term.

Importantly, all three terms are expressible through 
combinations of tree amplitudes and a simple kernel.
We have thus determined how general one-loop amplitudes transform under
the free superconformal symmetries corrected by the collinear deformations
discussed in \secref{sec:exacttree}.
The deformed transformation law was verified explicitly
for the example of one-loop MHV amplitudes 
in the dimensional reduction scheme
in \cite{Beisert:2010gn}. 
One can even contemplate deforming the superconformal representation 
further by these three terms to make amplitudes manifestly invariant.

\medskip

To continue to higher loops, a promising proposal has been made in \cite{Sever:2009aa}. 
It consists in adding the terms in \figref{fig:fullinvariance}g--i 
to the general transformation rule. 
The complete rule apparently respects unitarity in the sense that 
it appears to formally commute with taking cuts
(this might require to further add the contributions in \figref{fig:fullinvariance}c,f).
One may therefore expect that a unitarity-based construction of loop amplitudes will respect the rule. 
A practical problem is that the terms \figref{fig:fullinvariance}f--i
suffer from the same problems as \figref{fig:oneloopinvariance}c:
the subamplitude has to be evaluated right on a singularity. 
Here it does not suffice to discard the result, 
as it also contains important finite contributions. 
In \cite{Sever:2009aa} the terms are evaluated using the CSW rules 
formally leading to agreement.
We can also identify the three one-loop terms discussed above
in the various terms in the higher-loop rule:
The cusp anomalous dimension term in \figref{fig:oneloopinvariance}a,
the collinearity term in \figref{fig:oneloopinvariance}b
and the one-loop splitting term in \figref{fig:oneloopinvariance}c
represent cuts of the terms in 
\figref{fig:fullinvariance}f,g,h, respectively.

In conclusion, higher-loop superconformal transformations can be 
investigated by taking unitarity cuts. 
The rule in \figref{fig:fullinvariance} is promising, 
but its evaluation in practice is subtle. 
A two-loop analysis would be highly desirable to settle 
several open questions.

\section{Yangian Symmetry}
\label{sec:yangsym}

In the previous \Secref{sec:exact} we have discussed correction terms to
the free superconformal symmetry representation on scattering amplitudes in
$\superN=4$ SYM theory. Here we review how this Lie algebra symmetry
extends to an integrable structure in the planar limit.

\begin{figure}
\begin{center}
\includegraphics{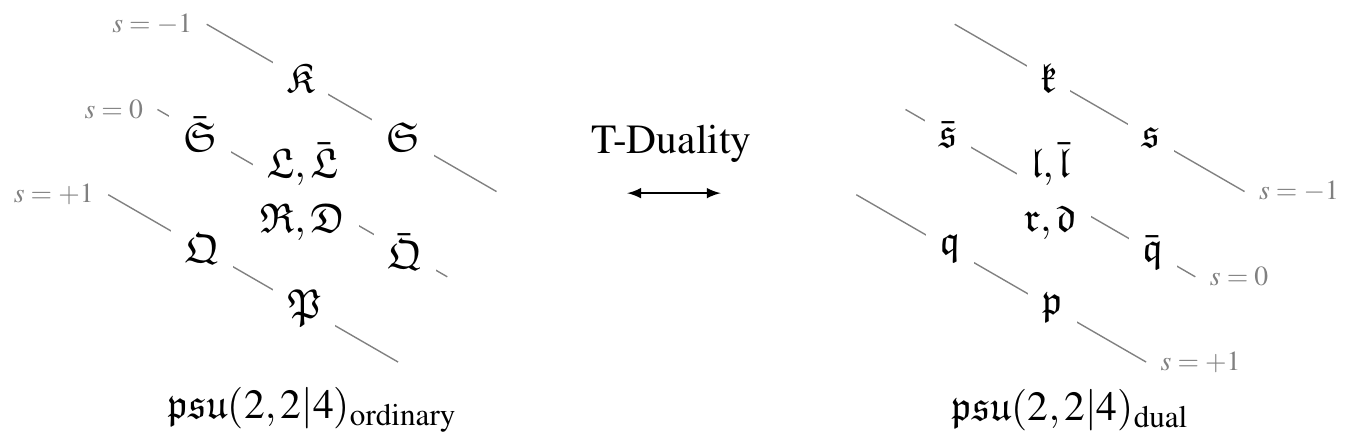}
\end{center}
\caption{The ordinary representation of the superconformal algebra and its
dual symmetry. Lines of constant eigenvalue under the commutator
with the sum of dilatation generator and hypercharge of $\alg{pu}(2,2|4)$
are indicated in gray. }
\label{fig:SymAmpWils}
\end{figure}

One of the most fundamental properties of the AdS/CFT system is that the
underlying symmetry of both $\superN=4$ SYM theory as well as type IIB
string theory on $\mathrm{AdS}_5\times\mathrm{S^5}$ is given by the
superconformal algebra $\alg{psu}(2,2|4)$. This symmetry is realized on
different observables on the two sides of the duality by respective
representations of the superconformal generators. 
In addition to this Lie algebra symmetry, there is a T-duality that leaves
the bulk action invariant and thus maps the string theory onto itself
\cite{Berkovits:2008ic,Beisert:2008iq}.
As a consequence, one may study the action of this T-self-duality on the
representation of the Lie algebra symmetry for different observables.
On the gauge theory side, however, a counterpart to the string T-duality is
not known. This fact obscures the explicit investigation of
the dual image of
the Lie symmetry representation on scattering amplitudes in $\superN=4$ SYM
theory, which is given in terms of generators
$\gen{J}_\alpha\in\{\gen{P},\gen{Q},\gen{\bar Q},\gen{D},\gen{L},\gen{\bar L},\gen{R},\gen{\bar S},\gen{S},\gen{K}\}$.%
\footnote{Here $\alpha$ labels the different generators of $\alg{psu}(2,2|4)$.}
Remarkably though, 
planar gauge theory amplitudes reveal a second
$\alg{psu}(2,2|4)$
Lie symmetry
represented by $\gen{j}_\alpha\in \{\gen{p},\gen{q},\gen{\bar
q},\gen{d},\gen{l},\gen{\bar l},\gen{r},\gen{\bar s},\gen{s},\gen{k}\}$
\cite{Drummond:2008vq}. Via the AdS/CFT duality, this second, so-called dual, 
superconformal symmetry is interpreted as the image of the ordinary
superconformal symmetry under the string theory T-duality, cf.\
\Figref{fig:SymAmpWils}. While the first symmetry corresponds to the
ordinary superconformal symmetry of scattering amplitudes, its dual image
can be understood as the ordinary symmetry of Wilson loops in $\superN=4$
SYM theory.%
\footnote{In fact, gauge theory scattering amplitudes and Wilson loops can
also be shown to map to each other
\cite{Alday:2007hr,Drummond:2007aua,Brandhuber:2007yx,Drummond:2007cf,Drummond:2007au,Drummond:2008aq}
-- at least in the case of MHV amplitudes and bosonic Wilson loops (cf.\
\cite{Mason:2010yk,CaronHuot:2010ek,Bullimore:2011ni,Belitsky:2011zm} for the supersymmetric
extension).
See also \cite{Korchemski:2011..} within this special issue.}

The above implies that the Lie algebra generators $\gen{J}$ on scattering
amplitudes are supplemented by additional operators $\gen{j}$ that furnish
dual symmetries. 
The latter have various different roles:
The operators $\{\gen{\bar q},\gen{d},\gen{l},\gen{\bar l},\gen{r},\gen{\bar s}\}$
are identical to $\{\gen{\bar S},\gen{D},\gen{L},\gen{\bar L},\gen{R},\gen{\bar Q}\}$,
(T-duality maps this $\alg{su}(2)\times\alg{su}(2|4)$ subalgebra to itself).
The operators $\gen{q}$ and $\gen{p}$ are
trivial when evaluated on amplitudes.
None of the above thus implies new symmetries.
Only the operators $\gen{s}$ and $\gen{k}$
are unrelated to the $\gen{J}$'s,
implying that the closure of these two superconformal algebras is bigger.
To be more precise, it is a Yangian algebra \cite{Drummond:2009fd}
generated by an infinite tower of $\alg{psu}(2,2|4)$-like charges.
Since in these arguments the role of amplitudes and Wilson
loops as well as that of their symmetries is interchangeable, 
one can express the charges in either picture:
\begin{equation}%
\{\gen{J}=\gen{J}^{\levind{0}}, \,\,\geny{J}=\gen{J}^{\levind{1}},\,\,\gen{J}^{\levind{2}},\,\,\dots\}
\quad \simeq\quad
\{\gen{j}=\gen{j}^{\levind{0}},\,\, \geny{j}=\gen{j}^{\levind{1}},\,\,\gen{j}^{\levind{2}},\,\,\dots\}.
\end{equation}
Here we have chosen specific names for the first two levels since these are
sufficient to recursively generate the whole tower of generators via commutation, e.g.\
$\gen{J}^{\levind{2}}\simeq\comm{\geny{J}}{\geny{J}}$. The
algebra underlying each of these infinite sets of generators is the
so-called Yangian $\yangian{\psu}$ of the superconformal
group whose definition will be made precise below.
It consists of an infinite number of levels whose structure will be
explained in the next \Secref{sec:Yangianalgebra}.

\begin{figure}
\begin{center}
\includegraphics{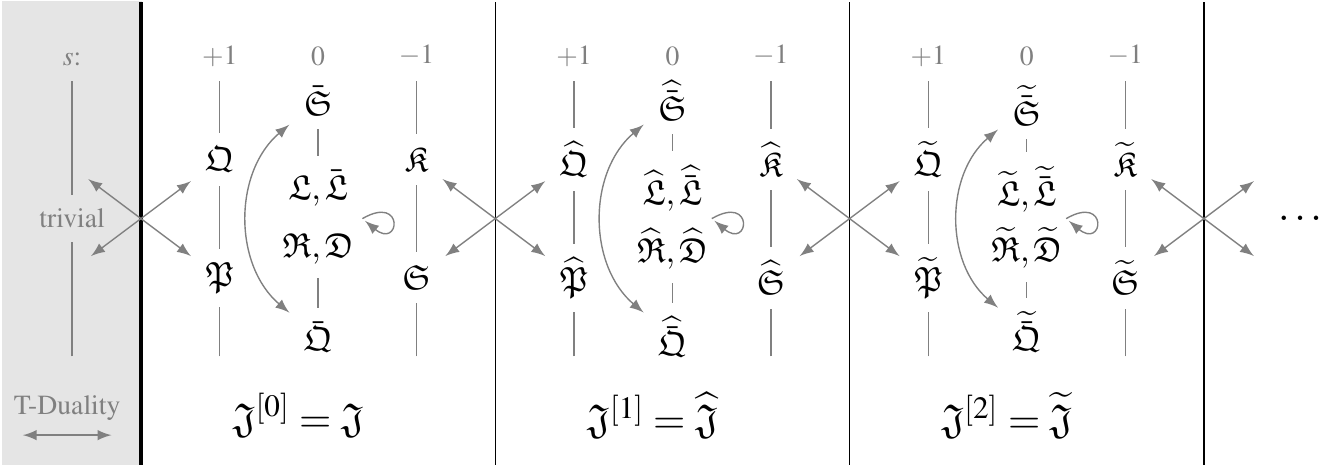}
\end{center}
\caption{The level-$r$ Yangian generators map under T-duality ($\to$) to
different levels according to their weight $s=\pm1,0$ under the sum of the dilatation
generator $\gen{D}$ and the hypercharge $\gen{B}$ of $\alg{pu}(2,2|4)$. }
\label{fig:TdualYangian}
\end{figure}

The T-duality can then be understood to map between the different levels of
the Yangian symmetry \cite{Beisert:2009cs}, cf. \Figref{fig:TdualYangian}:%
\footnote{Also here the role of $\gen{J}$ and $\gen{j}$ can be interchanged.}
\begin{equation}\label{eq:TdualYang}
\gen{j}^{\levind{r}}_\alpha\simeq\pm\gen{J}^{\levind{r+s(\alpha) }}_{-\alpha},
\qquad\qquad
\comm{\gen{D}+\gen{B}}{\gen{J}^{\levind{r}}_\alpha}=s(\alpha) \,\gen{J}^{\levind{r}}_\alpha.
\end{equation}
Here $\gen{D}$ and $\gen{B}$ are the dilatation generator and hypercharge
of $\alg{psu}(2,2|4)$, respectively.
Furthermore, for the index
$\alpha=[\gen{P},\gen{Q},\gen{\bar Q},\gen{D},\gen{L},\gen{\bar L},
\gen{R},\gen{\bar S},\gen{S},\gen{K}]$ 
we define the conjugate index 
$-\alpha=[\gen{K},\gen{S},\gen{\bar S},\gen{D},
\gen{L},\gen{\bar L},\gen{R},\gen{\bar Q},\gen{Q},\gen{P}]$
and the shift
$s(\alpha)=[1,1,0,0,0,0,0,0,-1,-1]$.
 
\subsection{Yangian Algebra}
\label{sec:Yangianalgebra}

The realization of Yangian symmetry on scattering amplitudes and Wilson
loops within the AdS/CFT duality gives an astonishing example of this
algebra. Its mathematical structure, however, can be formulated on a more
abstract level without any notion of dual symmetry:

A Yangian algebra $\yangian{\alg{g}}$ associated with a Lie algebra
$\alg{g}$ is defined by two sets of generators $\GG_\alpha$ and
$\widehat{\GG}_\beta$ obeying the following axioms \cite{Drinfeld:1985rx}:
\begin{enumerate}\itemsep0pt%
\refstepcounter{equation}\label{eq:Yangaxioms}
\def\theenumi{\roman{enumi})}\def\labelenumi{\theenumi}
\item\label{yang1}
\makebox[6.7cm][l]{Ordinary Lie Symmetry:} $\comm{\GG_\alpha}{\GG_\beta}=f_{\alpha\beta}{}^\gamma \GG_\gamma$,
\item\label{yang2}
\makebox[6.7cm][l]{Adjoint Level-One Symmetry:} $\comm{\GG_\alpha}{\widehat{\GG}_\beta}=f_{\alpha\beta}{}^\gamma \widehat{\GG}_\gamma$,
\item\label{yang3}
Serre-Relations:\qquad
$\comm{\widehat{\GG}_\alpha}{\comm{\widehat{\GG}_\beta}{\GG_\gamma}}+\text{two cyclic}= f_{\alpha\rho}{}^\lambda f_{\beta\sigma}{}^\mu f_{\gamma\tau}{}^\nu f^{\rho \sigma\tau}   \GG_{\{\lambda}\GG_\mu\GG_{\nu\}}$.
\hfill(\theequation)
\end{enumerate}
Here $f_{\alpha\beta\gamma}$ denotes the structure constants of the Lie
algebra $\alg{g}$ spanned by the generators $\GG_\alpha$,
and indices are raised by the Cartan--Killing form. 
The generalization to superalgebras is given by a straight-forward grading of
these relations. 

We will assume a specific (tensor product) representation for the level-zero,
i.e.\ the standard Lie symmetry generators $\GG_\alpha$ acting on a tensor
product of vector spaces $\mathbb{V}_k$ by (cf. \eqref{eq:Sbar0})
$\GG_\alpha=\sum_k \GG_{\alpha,k}$. Based on such a representation,
Drinfel'd introduced the following bilocal definition of additional
generators \cite{Drinfeld:1985rx}:
\begin{equation}\label{eq:levone}
\widehat{\GG}_\alpha=\sum_{1\leq\ell<k\leq n}f_\alpha{}^{\gamma\beta}\,\GG_{\beta,\ell}\,\GG_{\gamma,k}+
  \sum_{1\leq k\leq n} u_k\, \GG_{\alpha,k}.
\end{equation}
For many algebras and corresponding representations $\GG_\alpha$ --
including all known occurrences within the AdS/CFT correspondence -- the
definition \eqref{eq:levone} yields generators $\widehat{\GG}_\alpha$ that
indeed obey the axioms \eqref{eq:Yangaxioms} 
and thus generate a Yangian.%
\footnote{In order to prove the Serre-relations it suffices to show that
the right hand side of \ref{yang3} in \eqref{eq:Yangaxioms}
vanishes on one vector space
of the tensor product. This is due to the fact that the Yangian is a Hopf
algebra whose coproduct $\mathrm{\Delta}: \mathbb{V}\to \mathbb{V}\otimes\mathbb{V}$
is compatible with the Serre relations. 
(cf.\ \cite{Dolan:2004ps,Bargheer:2010hn} for different proofs of the
Serre-relations in the context of supersymmetric gauge theories).}
Furthermore, in most physical applications including our current one 
the evaluation parameters $u_k=u$ are all equal.%
\footnote{The value of $u$ does not make a difference as it merely 
multiplies the level-zero representation,
conventionally one sets $u=0$.}

Let us briefly comment on the relation of the Yangian to standard integrable models without going into details. 
Typically integrable systems can be based on a 
Lax operator obeying the Yang--Baxter equation (e.g.\
spin chains) used to define a monodromy matrix $\mathcal{M}(u)$ as a
function of the spectral parameter $u$. For an $\alg{su}(N)$ symmetric
model, the Yangian levels can then be interpreted as the different orders
in the expansion of this monodromy around the point where the Lax operator
reduces to the identity (conventionally $u=\infty$): 
\begin{equation}
\mathcal{M}(u=\infty)\simeq \mathbb{I}+\sum_{k=0}^\infty u^{-1-k} \GG^{\levind{k}}.
\end{equation}
Thus, the monodromy provides a way to determine the Yangian generators or,
turning the logic around, the representation of a Yangian algebra allows in
principle to reconstruct the monodromy of the integrable model. Such a
representation at hand, one may identify an integrable structure by showing
that the Yangian commutes with a theory's Hamiltonian 
(up to boundary terms)
or that its observables are invariant under this symmetry.

Color ordered scattering amplitudes in $\superN=4$ SYM theory are
\emph{cyclic} functions of the external particle degrees of freedom and
cyclicity is typically not compatible with Yangian symmetry. This is due to
the form of the Yangian level-one generator \eqref{eq:levone} given by an
ordered  sum over pairs of particles. In order to
investigate this problem, one can evaluate the
difference of two generators \eqref{eq:levone} shifted by one site
which corresponds to a one-site cyclic permutation of the amplitude's legs
\cite{Drummond:2009fd,Beisert:2010jq}:
\begin{equation}
\widehat{\GG}^\alpha\big[\raisebox{.15cm}{\text{\tiny $\sum\limits_{1\leq\ell<k\leq n}$}}\big]-\widehat{\GG}^\alpha\big[\raisebox{.15cm}{\text{\tiny $\sum\limits_{2\leq\ell<k\leq n+1}$}}\big]=\half f^\alpha{}_{\beta\gamma} f_\delta{}^{\beta\gamma}{}\,\,\GG_1^\delta
-f^\alpha{}_{\beta\gamma}\,\,\GG_1^\beta \GG^\gamma.
\label{eq:yangcyc}
\end{equation}
In general, this difference does not vanish. 
In the case of $\superN=4$ SYM theory amplitudes, however, there are two
further properties which lead to a well-defined Yangian despite of this
cyclicity:
\begin{enumerate}
\item The amplitudes transform as \emph{singlets} under the level-zero symmetry $\GG_\alpha$ of a Lie algebra $\alg{g}$.
\item The Lie algebra $\alg{g}$ has a \emph{vanishing} dual Coxeter number $f^\alpha{}_{\beta\gamma}f_\delta{}^{\beta\gamma}$.
\end{enumerate}
These two properties guarantee that the right hand side of
\eqref{eq:yangcyc} vanishes when evaluated on amplitudes and that in this
special case the Yangian level-one generators are compatible with
cyclicity. Consequently, the Yangian represents a well-defined symmetry of
the color ordered amplitudes.

There is an additional important property of $\alg{g}=\alg{psu}(2,2|4)$:
The algebra may be enhanced to $\alg{pu}(2,2|4)$ by an external
automorphism or so-called hypercharge, typically denoted by the generator $\gen{B}$. 
This generator measures the overall helicity of scattering amplitudes,
i.e.\ amplitudes are generically not invariant under it.
Remarkably, it can be shown that
scattering amplitudes of $\superN=4$ SYM theory are invariant under
the level-one generator $\geny{B}$ associated to $\gen{B}$ \cite{Beisert:2011pn}. 
This bilocal generator together with the ordinary superconformal symmetry 
yields all previously
known symmetries (e.g.\ the dual symmetry) of scattering amplitudes in
$\superN=4$ SYM theory.%
\footnote{%
The symmetry algebra $\alg{g}=\alg{osp}(6|4)$ of $\superN=6$
superconformal Chern--Simons theory (to be discussed in \secref{sec:abjm} below)
is not enhanceable by an external
automorphism. This reflects the fact that helicity is absent in the
three-dimensional theory. Understanding the algebraic difference between
the symmetries of $\superN=4$ SYM theory and $\superN=6$ SCS theory might
eventually resolve the problems to formulate a T-self-duality for the
string dual of the latter gauge theory and to put the discovered dual
symmetry in three dimensions on solid grounds. Note in this context that
the role of the generator $\alg{B}$ in $\superN=4$ SYM theory shows formal
similarities to the trace of the R-symmetry in $\superN=6$ SCS theory.}

\subsection{Dual versus Yangian Symmetry at Tree-Level}
\label{sec:dualvsYangian}

It is instructive to see the explicit relation between dual conformal and
Yangian level-one generators \cite{Drummond:2009fd}. Let us therefore
consider the example of the dual conformal boost. On the full superspace
spanned by ordinary supersymmetric spinor ($\lambda,\tilde\lambda,\eta$)
and dual ($x,\theta$) coordinates, it takes the form 
\begin{equation}\label{eq:dualconfboost}
\gen{k}^{a\dot a}
=\sum_{i=1}^n
\lrbrk{
x_i^{a\dot b}x_i^{\dot a b}\frac{\partial}{\partial x_i^{b\dot b}}
+x_i^{\dot a b}\theta_i^{aB}\frac{\partial}{\partial \theta_i^{\beta B}}
+x_i^{\dot a b}\lambda_i^a\frac{\partial}{\partial \lambda_i^\beta}
+x_{i+1}^{a\dot b}\tilde\lambda_i^{\dot a}\frac{\partial}{\partial \tilde \lambda_i^{\dot b}}
+\tilde \lambda_i^{\dot a}\theta_{i+1}^{a B} \frac{\partial}{\partial \eta_i^B}
}.
\end{equation}
Here the dual coordinates providing the natural variables of Wilson loops in $\superN=4$ SYM theory are defined by
\begin{equation}\label{eq:dualcoord}
x_i^{a\dot{a}}-x_{i+1}^{a\dot{a}}=\lambda_i^a\tilde\lambda_i^{\dot{a}}\,,
\qquad
\theta_i^{aA}-\theta_{i+1}^{aA}=\lambda_i^a\eta_i^A\,.
\end{equation}
The form of \eqref{eq:dualconfboost} on the full superspace results from
requiring that the generator commutes with the constraints
\eqref{eq:dualcoord}.
Scattering amplitudes transform covariantly under the action of the dual conformal boost
generator, i.e.\ $\gen{k}^{a\dot a}A_n=-\sum_{i=1}^nx_i^{a\dot a}A_n$, which
motivates the redefinition
\begin{equation}
\gen{\tilde k}^{a \dot a}=\gen{k}^{a\dot a}+\sum_{i=1}^nx_i^{a\dot a}.
\end{equation}
Acting on amplitudes such that
we can neglect terms annihilating $A_n$, this
operator can be rewritten as the level-one Yangian generator
$\geny{P}^{a\dot a}$, whose form follows from the definition
\eqref{eq:levone}:
\begin{equation}\label{eq:p1gen}
\left.\gen{\tilde k}^{a \dot a}\right|_{A_n}=
\geny{P}^{a\dot a}=
\sum_{\ell<k}
\big[
\gen{P}_{\ell,c\dot c}
(\gen{L}_{k,a}^c\delta_{\dot a}^{\dot c}
+\gen{\bar L}_{k,\dot a}^{\dot c}\delta_a^c-\gen{D}_k\delta_a^c\delta_{\dot a}^{\dot c})
-\gen{Q}_{\ell,a}^C\gen{\bar Q}_{k,{\dot aC}}
-(\ell\leftrightarrow k)
\big].
\end{equation}
An analogous relation holds for $ \gen{ s}_{a}^A$ and
$\geny{Q}_a^A$ while all other dual conformal generators can be
related to the level-zero symmetry. 
Note that the bilocality in \eqref{eq:dualconfboost} is hidden in the
definition of the dual variables \eqref{eq:dualcoord}. In the Wilson loop
picture, $\gen{k}$ reduces to the ordinary conformal boost in coordinates
$x$ and $\theta$ and thus to the level-zero dual symmetry.

Invariance of tree-level scattering
amplitudes in $\superN=4$ SYM theory under Yangian symmetry can then be
seen in two ways: On the one hand, tree-level amplitudes can be written in
terms of manifestly dual superconformal invariant expressions making this
property obvious with regard to the above relations, cf.\ \Secref{sec:invagra}.
On the other hand one may in principle explicitly apply the simplest
level-one generator $\geny{P}$ as given in \eqref{eq:p1gen} to the
amplitudes and show invariance as done in \cite{Beisert:2010jq} for the MHV
case. The adjoint property 
\ref{yang2} of the Yangian \eqref{eq:Yangaxioms}
then guarantees invariance under the full algebra.

\subsection{Corrections to Yangian Generators}

As discussed in the previous sections, symmetry generators acting on
scattering amplitudes in $\superN=4$ SYM theory are affected by singularities.
These require corrections to the generators in order to render the symmetry
exact. The correction terms have to take into account the holomorphic
anomaly starting at tree-level as well as infrared singularities starting
at one-loop order. They also affect the level-one Yangian symmetry as will
be indicated here:

Two collinear massless particles are not distinguishable in a conformal
theory. At tree-level, this manifests itself in the occurrence of collinear singularities of the amplitudes
which violate their invariance under the free conformal symmetry.
As a consequence, the conformal generators $\gen{S}$, $\gen{\bar S}$
and $\gen{K}$ of the ordinary superconformal symmetry acquire
correction terms on the subspace of two-particle collinearities as shown
above \cite{Bargheer:2009qu}. 
At tree-level, these are the only correction
terms of the level-zero generators. In particular, the tree-level generator
of anomalous dimensions $\gen{D}$ does not obtain corrections. The
level-one symmetry inherits the correction terms from the conformal
level-zero generators via its bilocal definition \eqref{eq:levone}. 
This allows to explicitly determine all level-one corrections at tree level. As an example, the level-one generator $\geny{P}$
\eqref{eq:p1gen} obtains no tree-level correction since it does not depend
on $\gen{S}$, $\gen{\bar S}$ or $\gen{K}$. 

Then, for instance, the level-one tree-level correction $\geny{Q}_+$ to $\geny{Q}$, can be written as a
commutator of the form
\begin{equation}
\delta_{\dot b}^{\dot a}\,\geny{Q}_+^{aA}=\comm{\geny{P}^{a\dot a}}{\gen{\bar S}^A_{+,\dot b}}.
\end{equation}
Provided the
adjoint property \ref{yang2}
of the Yangian can be proved, all other level-one generators --
including their corrections -- could be obtained by commutation of
$\geny{P}$ with the level-zero symmetry. 
Note that this generically yields bilocal operators that change the number
of external particles of the amplitude.

At loop order, conformal symmetry is typically broken by the
renormalization scheme, e.g.\ by dimensional regularization which
introduces a mass scale $\mu$ and a regularization parameter $\epsilon$. In
order to render conformal symmetry exact, these parameters can be included
into correction terms to the level-zero symmetry. At one loop order, all
four conformal generators $(\gen{S},\gen{\bar S},\gen{K},\gen{D})$ obtain such corrections 
as demonstrated
in \Secref{sec:exact} \cite{Beisert:2010gn}. 
The loop corrections to the level-one symmetry require local terms 
reminiscent of those multiplied by the $u_k$'s in \eqref{eq:levone}.
At one loop order they take the perturbative form%
\footnote{As the corrections do not act on single legs
there is no canonical prescription for the summation bounds.
The local term $\widehat{\GG}_{\alpha,k}^{(1)}$
thus depends on the prescription and specifies the action at the bounds.}
\begin{equation}
\widehat{\GG}_\alpha^{(1)}
=\sum_{1\leq\ell<k\leq n}f_\alpha{}^{\gamma\beta}\,
\lrbrk{
\GG_{\beta,\ell}^{(1)}\,\GG_{\gamma,k}^{(0)}
+\GG_{\beta,\ell}^{(0)}\,\GG_{\gamma,k}^{(1)}
}
+\sum_{1\leq k\leq n} \widehat{\GG}_{\alpha,k}^{(1)}.
\end{equation} 
Let us again consider the
simplest level-one generator $\geny{P}^{(1)}$. Its
form can be obtained by acting with $\geny{P}^{(0)}$ onto the
one-loop amplitude and requiring invariance
$\geny{P}^{(0)}A^{(1)}+\geny{P}^{(1)}A^{(0)}=0$
\cite{Beisert:2010gn}:
\begin{equation}
\bigbrk{\geny{P}^{(1)}}^{a \dot a}
=
\sum_{1\leq\ell<k\leq n}
\bigl[
\gen{D}^{(1)}_{\ell,\ell+1}\gen{P}_k^{a \dot a} 
-\gen{P}_\ell^{a \dot a} \gen{D}^{(1)}_{k-1,k}
\bigr].
\end{equation}
Here the nontrivial contribution comes from the one-loop correction to the
dilatation generator \eqref{eq:corrD1}.
In fact, it is well-known that the dual conformal boost $\gen{\tilde k}$
alias $\geny{P}$ is anomalous at loop level
\cite{Drummond:2007au,Drummond:2008vq,Brandhuber:2009kh}. The conjectured
all loop form for the former allows to derive a similar expression for its
Yangian level-one counterpart in analogy to \eqref{eq:corrD1} \cite{Beisert:2010gn}:
\begin{equation}
\geny{P}(\lambdaYM)^{a\dot a}=(\geny{P}^{(0)})^{a\dot a}+ \Gamma(\lambdaYM,\epsilon) (\geny{P}^{(1)})^{a\dot a}.
\end{equation}
This equation including
\eqref{eq:frontcoeff}
is conjectured to guarantee invariance to all loop orders.

The most urgent question concerning the corrected Yangian generators is
whether and how the axioms 
\ref{yang1}, \ref{yang2} and \ref{yang3} in \eqref{eq:Yangaxioms} 
are compatible with the correction
terms to the generators.
In \Secref{sec:furthcon} it was already
indicated that at tree-level the corrections to the Lie algebra symmetry
modify axiom \ref{yang1} 
by gauge transformations.

\section{Invariants \& Gra{\ss}mannian}
\label{sec:invagra}

In integrable models, physical quantities are commonly severely constrained
or even fully determined by the enlarged symmetry. Thus one may hope that
the Yangian symmetry allows to express all \sym\ scattering amplitudes
in terms of a finite set of algebraic, differential, or integral
equations. In this section, we review the Yangian invariance
properties of scattering amplitudes, and comment on the implications of the
deformation. The presentation mostly focuses on
the tree-level case.

\subsection{Free Invariants: The Gra{\ss}mannian Formula}
\label{sec:freeinv}

In the following, we discuss invariants of the free, undeformed symmetries.
These are not exact invariants (see \secref{sec:exact}), but we ignore this
fact for the moment, and discuss the deformation and its exact invariants
in \secref{sec:exinv} below.
Tree-level amplitudes
$A_{n,k+2}$ are linear combinations of terms
$R_{n,k,a}$
which are individually
(almost) invariant under both ordinary and dual superconformal symmetry,
and hence also under Yangian
symmetry
\cite{Drummond:2008vq,Brandhuber:2008pf,Drummond:2008bq,Drummond:2008cr}
\begin{equation}
\amp_{n,k+2}=\amp_n\supup{MHV}\sum\nolimits_{a}R_{n,k,a}\,.
\end{equation}
Here, $k$ specifies the degree $4k$ of polynomials
in the fermionic variables or equivalently the helicity $h=n-2(k+2)$ of the amplitude.
The MHV prefactor
$\amp_n\supup{MHV}$ is Yangian
invariant by itself.
The tree-level dual superconformal invariants $R_{n,k,a}$ were constructed
recursively in \cite{Drummond:2008cr}.%
\footnote{See also \cite{Drummond:2011..} within this special issue.}
A generating function for all these invariants was given in
\cite{ArkaniHamed:2009dn},
which takes a surprisingly compact form.%
\footnote{See also \cite{Adamo:2011pv,Drummond:2011..} within this special issue.}
It can be written as \cite{Mason:2009qx}%
\begin{equation}
\mathcal{R}_{n,k}(\gamma;\twi{W})=\int_{\gamma}\frac{\dd\nu(t)}{M_1\cdots M_n}\,\deltad{4k|4k}(t\cdot\twi{W})\,,
\label{eq:lms}
\end{equation}
where $t$ is a complex $k\times n$ matrix, the dot denotes matrix multiplication,
and $\twi{W}=(\twi{W}_1,\dots,\twi{W}_n)\transpose$ are momentum-twistor variables as
introduced in \cite{Hodges:2009hk}:
\begin{equation}
\twi{W}_i^{\twind{A}}=(\lambda_i^a,\mu_i^{\dot{a}},\chi_i^A)\,,
\qquad
\mu_i^{\dot{a}}=x_i^{a\dot{a}}\lambda_{ia}\,,
\qquad
\chi_i^A=\theta_i^{aA}\lambda_{ia}\,.
\end{equation}
These are the twistors associated to the dual variables (or \emph{region
momenta}) $x_i$, $\theta_i$ defined in \eqref{eq:dualcoord}.
The symbols $M_i$ in
the denominator denote minors of the matrix $t$ made of $k$ successive
columns, starting with column $i$. The integration measure $\dd\nu(t)$ was
given explicitly in \cite{Mason:2009qx}. It has degree $k(n-k)$, and turns
the function into a multi-dimensional complex contour integral. 
Different invariants $R_{n,k,a}=\mathcal{R}_{n,k}(\gamma_a)$ are generated by distinct
contours $\gamma_a$ encircling different residues. Including the measure, the
integrand is invariant under local $\grp{GL}(k)$
``gauge'' transformations
acting on the rows of the matrix $t$.%
\footnote{This is the reason for the degree of the naive integration
measure $\dd^{k\cdot n}t$ being reduced to $k(n-k)$; otherwise, the
integral would be ill-defined.}
The space being integrated over thus is the
\emph{Gra{\ss}mannian}
$\mathrm{Gr}(k,n)$ 
consisting of all $k$-planes within
$\Complex^n$, where each plane is spanned by the $k$ rows of $t$.
For practical purposes, a gauge can be fixed by setting $k^2$ components of
$t$ to specific values.
A convenient gauge 
fixes the first $k$ columns of $t$ to the identity matrix
\begin{equation}
t=(1|\cdot)\,,
\qquad
\dd\nu(t)=\prod_{a=1}^k\prod_{i=k+1}^n\dd t_{ai}\,.
\label{eq:grassgauge}
\end{equation}

The benefit of using momentum twistors is that the dual superconformal
generators are realized linearly in these variables,
\begin{equation}
\gen{j}^{\twind{A}}{}_{\twind{B}}=\sum_{i=1}^n\twi{W}_i^{\twind{A}}\pardel{\twi{W}_i^{\twind{B}}}\,.
\label{eq:levelzero}
\end{equation}
Invariance under these generators is ensured by the delta function in
\eqref{eq:lms}. It has been shown in \cite{Drummond:2010qh} that taking
these dual generators as the level-zero algebra results in the same Yangian as
taking the ordinary superconformal symmetry as level-zero generators, which
is a consequence of the T-self-duality mentioned in \secref{sec:yangsym}.
The level-one Yangian generators take the usual form \eqref{eq:levone} of bilocal
combinations of (dual) level-zero generators,
\begin{align}
\geny{j}^{\twind{A}}{}_{\twind{B}}
&=\biggbrk{\sum_{i<j}-\sum_{j<i}}(-1)^{\twind{C}}\,
	\twi{W}_i^{\twind{A}}\pardel{\twi{W}_i^{\twind{C}}}\twi{W}_j^{\twind{C}}\pardel{\twi{W}_j^{\twind{B}}}
\,.
\label{eq:levelonesums}
\end{align}
Closely following \cite{Drummond:2010qh}, we will now show that the
function \eqref{eq:lms} is indeed invariant under the
level-one generators and thus under the whole Yangian algebra.
It is sufficient to show invariance under the first sum in
\eqref{eq:levelonesums}, as invariance under the second sum is completely
analogous.
The first sum can be expressed as
\begin{equation}\label{eq:4.8}
\sum_{i<j}\biggbrk{
	\twi{W}_i^{\twind{A}}\pardel{\twi{W}_j^{\twind{B}}}\twi{W}_j^{\twind{C}}\pardel{\twi{W}_i^{\twind{C}}}
	-\twi{W}_i^{\twind{A}}\pardel{\twi{W}_i^{\twind{B}}}}\,.
\end{equation}
Due to the linearity of the delta function's argument, the twistorial
operators $\twi{W}_j^{\twind{C}}\,\partial/\partial\twi{W}_i^{\twind{C}}$
can be replaced by operators $O_{ij}$ acting on the integration variables
$t_{ai}$.%
\footnote{In the
gauge \eqref{eq:grassgauge},
the operators are
$O_{i,j\leq k}=-\sum_{l=k+1}^nt_{jl}\pardel{t_{il}}$ and
$O_{i,j>k}=\sum_{b=1}^kt_{bi}\pardel{t_{bj}}$. While the form of
$O_{ij}$ for $j>k$ is derived straightforwardly, one needs to make use of
the delta function constraints to arrive at the form for $j\leq k$.}
The action of the level-one generators on $\mathcal{R}_{n,k}$ becomes
\begin{equation}
\geny{j}^{\twind{A}}{}_{\twind{B}}\,\mathcal{R}_{n,k}
=\int\frac{\dd\nu(t)}{M_1\cdots M_n}
 \sum_{a=1}^k\bigbrk{O_a^\twind{A}-V_a^\twind{A}}
 \partial_{a\twind{B}}\deltad{4k|4k}(t\cdot\twi{W})\,,
\label{eq:O-V}
\end{equation}
where
\begin{equation}
O_a^{\twind{A}}=\sum_{i<j}\twi{W}_i^{\twind{A}}O_{ij}t_{aj}\,,
\qquad
V_a^{\twind{A}}=\sum_{i<j}\twi{W}_it_{ai}\,,
\label{eq:OV}
\end{equation}
and $\partial_{a\twind{B}}=\partial/\partial\twi{W}_a^{\twind{B}}$.
Making use of the triangular form of $O_a^{\twind{A}}$, one can show that
the
$V_a^\twind{A}$-term cancels when commuting $O_a^\twind{A}$ past the
minors, $\comm{1/M_1\cdots M_n}{O_a^{\twind{A}}}=V_a^{\twind{A}}/M_1\cdots
M_n$. Thus
\begin{align}
\geny{j}^{\twind{A}}{}_{\twind{B}}\,\mathcal{R}_{n,k}
&=\int\dd\nu(t)\sum_{a=1}^kO_a^{\twind{A}}\frac{1}{M_1\cdots M_n}
  \partial_{a\twind{B}}\deltad{4k|4k}(t\cdot\twi{W})\,.
\label{eq:Jtotderiv}
\end{align}
Now each term in the integrand is a total derivative of a single-valued
function in one of the integration variables,
hence the integral along any closed contour vanishes.%
\footnote{This argument relies on the integration measure being a
gauge-invariant generalization of the standard measure
\eqref{eq:grassgauge}
\cite{Mason:2009qx}.}
This shows that $\mathcal{R}_{n,k}$ is indeed Yangian invariant.

The function $\mathcal{R}_{n,k}$ \eqref{eq:lms} in fact produces all
Yangian invariant terms $R_{n,k,a}$ that, multiplied by the $n$-point
MHV amplitude, form the planar $n$-point N$^k$MHV
amplitude \cite{Bourjaily:2010kw}. It is equivalent \cite{ArkaniHamed:2009vw}
to the previously proposed \cite{ArkaniHamed:2009dn} generating function
$\mathcal{A}_{n,k}(\twi{Z})$, which generates planar
tree-level amplitudes including the MHV prefactor. Formally, the relation
between the two functions is simply
\begin{equation}
\mathcal{A}_{n,k}(\twi{Z})=\amp_n\supup{MHV}\mathcal{R}_{n,k}(\twi{W})\,,
\label{eq:lah}
\end{equation}
where $\twi{Z}_1,\dots,\twi{Z}_n$ are ordinary spacetime twistors
$\mathcal{Z}_i^{\mathcal{A}}=(\partial/\partial\lambda^{a}_i,\tilde\lambda^{\dot a}_i,\eta^A_i)$ 
as opposed to momentum twistors.%
\footnote{$\partial/\partial\lambda$ indicates a Fourier transform w.r.t.\ $\lambda$.
Formally this required that $\lambda$ and $\tilde\lambda$ are unrelated
as in split spacetime signature $(2,2)$. See \cite{Wolf:2010av} for a
pedagogical review of twistor theory in the context of scattering
amplitudes.}
The MHV amplitudes $\amp\supup{MHV}$ are
Yangian invariant on their own.

It has been argued that the Gra{\ss}mannian integral
$\mathcal{R}_{n,k}$ \eqref{eq:lms} in fact
generates \emph{all} invariants of the free Yangian symmetry
\cite{Drummond:2010uq,Korchemsky:2010ut}. Assuming that all invariants of
\psu\ in the representation \eqref{eq:levelzero} are of the form
$\deltad{4k|4k}(t\cdot W)$, the most general \psu\ invariant
is exactly given by $\mathcal{R}_{n,k}$, except for the integration measure
being generalized by an arbitrary function $f(t)$ of the integration
variables. Requiring invariance under the level-one generators
\eqref{eq:levelonesums}, constraints on the function $f(t)$ are derived in
\cite{Drummond:2010uq,Korchemsky:2010ut}. Under certain assumptions,
the only remaining solution is a constant.

\subsection{Exact Invariants}
\label{sec:exinv}

So far, we have discussed invariants of the free, undeformed Yangian
symmetry. Physical scattering amplitudes are linear combinations of these
free invariants. On their own, the free Yangian invariants have no local
interpretation. They have unphysical
`spurious' singularities, and a wrong behavior in collinear limits. While
the free Yangian symmetry determines amplitudes to a large extent, it puts no
constraints on the coefficients of the physical linear combination. On the other
hand, if \sym\ is an integrable theory, one would expect all dynamical quantities
to be completely determined by the extended symmetry. The deformations
introduced in \secref{sec:exact} exactly appear to provide the missing piece.
Namely, under mild assumptions, the coefficients of the physical linear
combination appear to be
uniquely fixed by requiring the correct behavior in collinear
limits (or, alternatively, the cancellation of all spurious poles) \cite{Korchemsky:2009hm}.%
\footnote{The authors of \cite{Korchemsky:2009hm} show that requiring
correct collinear limits is sufficient for determining NMHV amplitudes.}
As the interaction terms in the deformed superconformal and Yangian
generators impose
precisely the correct collinear behavior, it is plausible
that only the physical linear combinations form invariants of the full
(deformed) classical Yangian.

Of course, the correct coefficients for all tree-level amplitudes are known
explicitly \cite{Drummond:2008cr}. Nevertheless, the extent to which the
symmetries determine the amplitudes is an important question. In
particular, a unique invariant at tree level is essential for a complete algebraic
determination of loop-level amplitudes. Namely, tree-level invariants form
the space of homogeneous solutions to the invariance equations at loop
level. Thus, they can be freely added to loop-level invariants.%
\footnote{Adding the physical tree-level amplitude can be compensated by
rescaling the coupling constant and the overall coefficient, both of which
cannot be determined algebraically in any case.}
Hence, if there would be multiple tree-level invariants, loop-level
amplitudes could not be determined uniquely.

\subsection{Loop Level}

At loop level, infrared divergences obscure the symmetry properties of
scattering amplitudes. For instance, free Yangian symmetry is broken due to (dimensional)
regularization. However, loop
amplitudes in \sym\ are to a large extent determined by their
singularities. The higher their codimension, the less are these
singularities affected by infrared divergences. In particular, the
``leading singularities'' with maximal codimension localize all loop
integrals; they can be expressed
entirely in terms of tree-level amplitudes and do not require
regularization, which makes them especially accessible.
In fact, it is conjectured that the function $\mathcal{A}_{n,k}$
\eqref{eq:lah} besides
all tree-level amplitudes generates all leading
singularities to arbitrary loop order in planar perturbation theory,
which would show that
these are invariant under the free Yangian
\cite{ArkaniHamed:2009dn}.%
\footnote{Up to contributions from collinear momenta, of course.}
Recently, a recursive construction of the planar integrand to arbitrary loop order
was proposed based on the assumption of invariance under the free Yangian
\cite{ArkaniHamed:2010kv}. The resulting integral of course is infrared
divergent. For treating the divergences, a useful scheme is provided by the
mass regularization of \cite{Alday:2009zm}.%
\footnote{See also \cite{Henn:2011xk} within this special issue and references
therein.}

For the integrated (infrared divergent) amplitudes, exact Yangian symmetry can be restored at one-loop
order by adding appropriate corrections \cite{Beisert:2010gn} along the
lines of \secref{sec:exact}.

\section{Symmetries of ABJM Amplitudes}
\label{sec:abjm}

Recently, a superconformal gauge theory in three dimensions (\scs, ABJM) was found
\cite{Bagger:2006sk,Gustavsson:2007vu,Bagger:2007jr,Bagger:2007vi,Gustavsson:2008bf,Bagger:2008se,Aharony:2008ug}
which bears remarkable similarities to four-dimensional \sym. In
particular, its planar spectrum of local operators at weak coupling is described by an
integrable spin chain 
(see \cite{Klose:2010ki} for a review).
Compared to \sym, however, much less is known about scattering
amplitudes in its three-dimensional cousin. Nevertheless,
counterparts to some of the most important symmetry
structures known from \sym\ amplitudes have been found for the
three-dimensional theory during the last year.
First, the four- and six-point tree-level amplitudes of \scs\ were shown to be
invariant under a Yangian symmetry algebra \cite{Bargheer:2010hn}.
Subsequently, a Gra{\ss}mannian formula for all tree-level amplitudes
\cite{Lee:2010du} as well as a dual superconformal symmetry
\cite{Huang:2010qy} were proposed.
On-shell recursion relations \`a la BCFW
\cite{Britto:2004ap,Britto:2005fq} for all \scs\ tree-level amplitudes were
presented in \cite{Gang:2010gy}, and were used to inductively demonstrate their
dual superconformal alias Yangian invariance.

Also in this theory, amplitudes can be formulated in terms of a superfield
\begin{equation}
\Phi=\phi^4+\eta^A\psi_A+\half\eps_{ABC}\eta^A\eta^B\phi^C+\sfrac{1}{6}\eps_{ABC}\eta^A\eta^B\eta^C\psi_4\,,
\label{eq:N6superfield}
\end{equation}
which, together with its conjugate $\bar\Phi$, captures all on-shell
dynamical degrees of freedom (eight scalars $\phi^A,\bar\phi_A$ and eight
fermions $\psi_A,\bar\psi^A$). The superconformal algebra \osp\
in three dimensions is realized in terms of the fermionic $\alg{u}(3)$
spinor $\eta^A$ and the real two-component spacetime spinor $\lambda^a$,
which parametrizes a three-dimensional momentum as
$p^{ab}=\lambda^a\lambda^b$
\cite{Gunaydin:1984vz,Zwiebel:2009vb,Minahan:2009te,Papathanasiou:2009en}.
On scattering amplitudes $\amp(\Phi_1,\ldots,\Phi_n)$, the superconformal
generators $\gen{J}_{\alpha}\in\osp$ act locally, while the Yangian level-one generators
$\geny{J}_{\alpha}$ take the usual bilocal form \eqref{eq:levone}:%
\footnote{This definition is compatible with the cyclicity of scattering
amplitudes because the dual Coxeter number of \osp\ vanishes, see also
\eqref{eq:yangcyc} above.}
\begin{equation}
\gen{J}_\alpha=\sum_{1\leq k\leq n}\gen{J}_{\alpha,k}\,,
\qquad
\geny{J}_\alpha=f_\alpha{}^{\gamma\beta}\sum_{1\leq j<k\leq n}\gen{J}_{\beta,j}\,\gen{J}_{\gamma,k}\,.
\end{equation}
Here, the generator $\gen{J}_k$ acts only on the coordinates of the $k$'th
leg $\Phi_k$. Interestingly, the R-symmetry is broken by the superfield
\eqref{eq:N6superfield} to a manifest
$\alg{u}(3)$ and a non-manifest remainder:
\begin{equation}
\gen{R}^{AB}=\eta^A\eta^B\,,
\qquad
\gen{R}^A{}_B=\eta^A\pardel{\eta^B}\,,
\qquad
\gen{R}_{AB}=\pardel{\eta^A}\pardel{\eta^B}\,.
\label{eq:N6R}
\end{equation}
In particular, the $\alg{u}(3)$
contains a non-vanishing trace
$\gen{R}^C{}_C=\eta^C\,\partial/\partial\eta^C-3/2$, which enforces
scattering amplitudes to be of homogeneous degree $\amp_n\sim(\eta)^{3n/2}$
in the fermionic variables. This implies
that there are no
``MHV-like'' amplitudes with a minimal degree in the fermionic variables.
The four- and six-point tree-level amplitudes have been computed%
\footnote{Four-point amplitudes of the mass-deformed theory had been
studied before in \cite{Agarwal:2008pu}.}
and shown to be invariant under the level-one momentum generator $\geny{P}$
\cite{Bargheer:2010hn}.
Invariance under all other
Yangian generators follows by commutation with level-zero generators.
For consistency of the Yangian, the Serre relations \eqref{eq:Yangaxioms}
have to be satisfied. While difficult to show in general, a rather direct
proof \cite{Bargheer:2010hn} utilizes the fact that the level-zero generators form a singleton
representation (as in four dimensions) that can be formulated in terms of
spinor-helicity variables.

Scattering amplitudes for higher numbers of legs are hard to compute, even
at tree level. However, a generating function for all \scs\ tree-level
amplitudes similar to \eqref{eq:lms} has been proposed in \cite{Lee:2010du}.
In spinor-helicity variables $\Lambda=(\lambda,\eta)$, it takes the form
\begin{equation}
\mathcal{A}_{2k}(\gamma;\Lambda)=\int_\gamma\frac{\dd\nu(t)}{M_1\cdots M_k}\,
	\deltad{k(k+1)/2}(t\cdot t\transpose)\,\deltad{2k|3k}(t\cdot\Lambda)\,.
\label{eq:grass3d}
\end{equation}
Here, $t$ is a $(k\times2k)$ matrix,
and the minors $M_j$ are defined as before. In
the four-dimensional case \eqref{eq:lms}, the domain of integration
was the Gra{\ss}mannian $\mathrm{Gr}(k,n)$, the space of all $k$-planes in
$\Complex^n$. Here, the additional delta function enforces the scalar
product to vanish on the $k$-plane spanned by the rows of $t$, which
restricts the domain of integration to the 
\emph{orthogonal Gra{\ss}mannian} 
$\mathrm{OGr}(k,2k)$, 
see \cite{Gang:2010gy}.
Again, all terms contributing to the
$2k$-point tree-level superamplitude are conjectured to be generated by
$\mathcal{A}_{2k}$ evaluated on different integration contours $\gamma$.
This has been verified for the four-point \cite{Lee:2010du} and the
six-point amplitude \cite{Gang:2010gy}. Moreover, the integral
\eqref{eq:grass3d} is Yangian invariant \cite{Lee:2010du}, which, assuming
Yangian symmetry for scattering amplitudes, is a strong hint for its
correctness.

The discovery of Yangian symmetry made the authors of \cite{Huang:2010qy}
formulate a dual superconformal symmetry for \scs\ amplitudes, as found
earlier for \sym.
By going to dual variables $x_j^{ab}$ with
\begin{equation}
\lambda_j^a\lambda_j^b=x_j^{ab}-x_{j+1}^{ab}
\label{eq:x}
\end{equation}
exactly as in four
dimensions, the proposed dual conformal symmetry (no super yet) acts on the dual variables
$x^{ab}$ in the same way the ordinary conformal symmetry acts on spacetime.
As all amplitudes only depend on differences of dual
$x_j$ variables, they are trivially invariant under dual translations
$\gen{p}_{ab}=\sum_j\partial/\partial x_j^{ab}$. Provided the
scattering amplitudes scale as
\begin{equation}
\amp_{2k}
\xrightarrow{\;\;\grp{I}\supup{dual}\;\;}
\sqrt{\prod\nolimits_{j=1}^{2k}x_j^2}\,\amp_{2k}\,,
\label{eq:inversionscaling}
\end{equation}
under dual inversions $\grp{I}\supup{dual}$, 
they also transform
covariantly under dual special conformal transformations $\gen{k}$,
\begin{equation}
\gen{k}_{ab}\,\amp_{2k}
=\grp{I}\supup{dual}\gen{p}_{ab}\grp{I}\supup{dual}\,\amp_{2k}
=-\frac{1}{2}\biggbrk{\sum_{j=1}^{2k}x_{j,ab}}\amp_{2k}\,.
\end{equation}
The dual conformal symmetry algebra is completed by Lorentz generators
$\gen{l}=\gen{L}$ and the dilatation generator $\gen{d}=\gen{D}$, which are
equal to the corresponding generators of the ordinary conformal symmetry.

Trying to extend the dual conformal to dual \emph{super}conformal symmetry,
one encounters an important difference to the four-dimensional case.
Namely, besides the fermionic variables $\theta_j^{aA}$ as known from \sym,
another set of dual variables $y_j^{AB}$ is required for formulating the
full dual symmetry. Here,
\begin{equation}
\lambda_j^a\eta_j^A=\theta_j^{aA}-\theta_{j+1}^{aA}\,,
\qquad
\eta_j^A\eta_j^B=y_j^{AB}-y_{j+1}^{AB}\,.
\label{eq:theta-y}
\end{equation}

Specifically, it is impossible to consistently express the action of some
of the dual generators on the original variables $(\lambda,\eta)$ without
also using the additional variables $y$.%
\footnote{More precisely, the dual generators cannot be formulated on the
``full space'' of \emph{independent} variables $(\lambda,\eta,x,\theta)$
while preserving the hypersurface constraints \eqref{eq:x,eq:theta-y}
without also using the additional variables $y$. This formulation is needed
though for finding the action of the dual generators on the original
variables $(\lambda,\eta)$, and for studying their relation to the ordinary
symmetry generators.}
The presence of a dual superconformal symmetry hints at a scattering
amplitude\,/\,Wilson loop duality like in \sym. Light-like Wilson loops
were studied and successfully compared to the tree-level%
\footnote{The one-loop contributions vanish in both cases.}
four-point scattering amplitude in \cite{Henn:2010ps}.%
\footnote{Very recently, also $n$-point correlation functions were related
to polygonal Wilson loops \cite{Bianchi:2011rn}.}

As in four dimensions, some of the dual generators $\gen{j}$ are trivial,
others are identical to their ordinary-symmetry counterparts, and some are
equal to level-one Yangian generators $\geny{J}$ when acting on invariants of
the ordinary conformal symmetry \cite{Huang:2010qy}:
\begin{align}
  (\gen{p}_{ab},\gen{q}_{ab},\gen{r}_{AB})
&=\text{trivial}\,,\nn\\
  (\gen{l}^a{}_b,\gen{d},\gen{r}^A{}_B,\gen{q}^A_a,\gen{s}^a_A)
&=(\gen{L}^a{}_b,\gen{D},\gen{R}^A{}_B,\gen{S}^A_a,\gen{Q}^a_A)\,,\nn\\
  ( \gen{k}^{ab}, \gen{s}^{aA}, \gen{r}^{AB})
&\simeq(\geny{P}^{ab},\geny{Q}^{aA},\geny{R}^{AB})\,.
\label{eq:dualstruct}
\end{align}
Invariance under the full dual superconformal symmetry thus follows from
invariance under the ordinary symmetry and under the dual generator
$\gen{k}\simeq\geny{P}$, for instance. Furthermore, the dual and the ordinary symmetry together
generate the whole Yangian algebra $\yangian{\osp}$.

In recent years, a key tool for the investigation of scattering amplitudes
in four dimensions have been the on-shell `BCFW' recursion relations
\cite{Britto:2004ap,Britto:2005fq}.%
\footnote{See also the review \cite{Wolf:2010av},
as well as \cite{Brandhuber:2011ke} within this special issue.}
A few months
ago, similar relations were found for \scs\ scattering amplitudes in three
dimensions \cite{Gang:2010gy}. Unlike their four-dimensional counterpart, the
three-dimensional recursion relations require shifting two external momenta
non-linearly in the auxiliary complex variable $z$. Namely,
\begin{align}
\lambda_j&\to+\half(z+1/z)\lambda_j+\ihalf(z-1/z)\lambda_k\,,\\
\lambda_k&\to-\ihalf(z-1/z)\lambda_j+\half(z+1/z)\lambda_k\,.
\end{align}
Using the recursion relations, the scaling \eqref{eq:inversionscaling}
under dual inversions was proved inductively, thus establishing dual
superconformal alias Yangian invariance for all tree-level amplitudes.
Furthermore, the amplitudes obtained by recursion were successfully matched
against the Gra{\ss}mannian formula \eqref{eq:grass3d} for up to eight
external particles.

The dual superconformal symmetry is particularly surprising because to date
no supersymmetric T-self-duality of the AdS/CFT dual sigma model
\cite{Arutyunov:2008if,Stefanski:2008ik} has been
found. In the case of \sym, dualizing the coordinates along the
$\gen{P}^{ab}$ and $\gen{Q}^{aB}$ directions of the supercoset
$\grp{PSU}(2,2|4)/\grp{Sp}(1,1)\times\grp{Sp}(2)$ maps the sigma model onto
itself, while turning ordinary into dual symmetry generators.
In contrast, it appears impossible to supersymmetrically extend
a bosonic T-duality involving only the translational directions of
$\grp{AdS}_4$ within the supercoset
$\grp{OSp}(6|4)/\grp{U}(3)\times\grp{SO}(3,1)$
\cite{Adam:2009kt,Grassi:2009yj}.
The sigma model on this coset is obtained by a kappa-gauge fixing that
is not compatible with all string configurations \cite{Gomis:2008jt}, and
it was suspected \cite{Adam:2010hh} that the gauge fixing could obstruct a
T-self-duality and/or dual symmetry
that might be present in the string theory.
But even using the full superspace formulation of \cite{Gomis:2008jt},
the extension of a pure $\grp{AdS}_4$ T-duality to a full 
self-duality appears
impossible \cite{Grassi:2009yj}.
A resolution could be to also T-dualize
some of the coordinates from the $\cp^3$ part of the bosonic background.
The structure of both the R-symmetry realization \eqref{eq:N6R} and the dual
symmetry \eqref{eq:dualstruct} suggests to dualize the coordinates along
the $\gen{R}^{AB}$ directions, which generate three Abelian isometries of
$\cp^3$. This has been attempted in \cite{Adam:2010hh}, but leads to a
singular transformation that could not be regularized thus far
\cite{Bakhmatov:2010fp} (see also \cite{Dekel:2011qw}).

Just as in four dimensions, one would expect the superconformal symmetry generators for
scattering amplitudes in \scs\ to receive corrections that have
distributional support on collinear momentum configurations. However, no
source for anomalous contributions from the free generators has been found
thus far.

\section{Summary \& Outlook}

Conformal symmetry implies powerful constraints on a physical theory.
Nevertheless, it is not always easy to implement it in a mathematically
concise way and the difficulties in finding an adequate representation may
be misinterpreted as a breaking of this symmetry. In the first part of this
review we have investigated two instances of this problem arising in
$\superN=4$ SYM theory:
\begin{itemize}
\item 
The holomorphic anomaly leads to a violation of the free superconformal
symmetry that can be overcome by corrections to the symmetry generators.
This modification only affects collinear momentum configurations and may
thus be associated with the ambiguous description of asymptotic states
for massless particles (starting at tree-level).

\item A renormalization scheme that introduces a mass scale superficially
breaks conformal symmetry. Also this shortcoming can be cured by adapting
the symmetry representation to the corresponding scheme (starting at
one-loop order).
\end{itemize}
The resulting corrections to the representation of the superconformal
algebra relate amplitudes for different numbers of particles and thereby
induce recursive relations among them. The representation obeys the
commutator relations modulo gauge transformations that vanish when
evaluated on scattering amplitudes.

In the second part of the review we have indicated how dual superconformal
symmetry
results in a Yangian algebra realized
on scattering amplitudes. This Yangian algebra forms a typical mathematical
structure underlying integrable models. Its representation inherits the
deformations of the Lie algebra symmetry mentioned above. While the free
representation is able to distinguish certain symmetry invariant building
blocks for the amplitudes, the deformation of the integrable structure is
crucial for fixing their exact linear combination. Importantly, the
building blocks can be generated by a Gra{\ss}mannian function and we have
commented on its relation to the Yangian symmetry.

Finally, similar observations made in $\superN=6$ SCS theory were
summarized. While their investigation is still in its infancy, there are
strong indications for Yangian symmetry, dual superconformal symmetry as
well as a Gra{\ss}mannian function paralleling the discoveries in
$\superN=4$ SYM theory. 

Several interesting problems arise in this context. Firstly, it would be
important to determine the conformally exact representation of
$\alg{psu}(2,2|4)$ at higher loop orders and to investigate the imposed
constraints on scattering amplitudes. This could reveal the full power of
the underlying formalism potentially facilitating the computation of so far
undetermined amplitudes. It would furthermore be crucial to verify the
Yangian algebra relations for the deformed representation at tree and loop
level. Only this would a posteriori justify the name \emph{Yangian} for the
discovered mathematical structure. It would then be highly desirable to
study the action of the deformed Yangian symmetry on the Gra{\ss}mannian
function in order to determine the impact of the correction terms. This
might yield a prescription for how to obtain the full superamplitude from a
generating function. Moreover it would be very interesting to construct
Yangian invariants from scratch, i.e.\ to study the invariants of the
corrected algebra as well as their uniqueness starting from the given
representation. The way in which the correction terms relate to the
Gra{\ss}mannian formulas described above might be extremely enlightening.
Very recently a new bilocal generator $\geny{B}$ corresponding to
the hypercharge of the superconformal algebra has been shown to annihilate
the amplitudes \cite{Beisert:2011pn}. It would be important to find out how
this generator can be embedded into the above context. Finally many of the
above problems carry over to the scattering problem of $\superN=6$ SCS
theory. Here the most pushing question is the relation of the discovered
algebraic symmetries to a potential T-duality of the
$\mathrm{AdS}_4\times\mathrm{\mathbb{C}P^3}$ superstring theory.

\pdfbookmark[2]{Acknowledgments}{acknowledgments}
\subsection*{Acknowledgments}

The work of TB was supported by the Swedish Research Council (VR) under
grant 621-2007-4177.
The work of NB is supported in part by 
the German-Israeli Foundation (GIF).

\pdfbookmark[1]{\refname}{references}
\bibliographystyle{nb}
\bibliography{ampreview}

\end{document}